\documentclass[twocolumn,preprintnumbers,amsmath,amssymb]{revtex4}

\usepackage{graphicx}% Include figure files
\usepackage{dcolumn}% Align table columns on decimal point

\def\nslash{n\!\!\!\slash}
\def\bnslash{\bar n\!\!\!\slash}

\def\epslash{\epsilon\!\!\!\slash}

\def\OMIT#1{}

\newcommand{\bn}{{\bar n}}
\newcommand{\bea}{\begin{eqnarray}}
\newcommand{\eea}{\end{eqnarray}}

\newcommand{\tdot}{\!\cdot\!}

\newcommand{\cC}{{\mathrm C}}
\newcommand{\cO}{{\mathcal O}}

\newcommand{\qbar}{{\bar{q}}}
\newcommand{\nbar}{{\bar{n}}}

\newcommand{\tmop}{\mathrm}

\begin{document}

\title{Improving jet distributions with effective field theory}

\author{Christian W.~Bauer}
\email{cwbauer@lbl.gov}
\author{Matthew D.~Schwartz}
\email{mdschwartz@lbl.gov} 
\affiliation{Ernest Orlando Lawrence Berkeley National Laboratory and
University of California, Berkeley, CA 94720}

\date{\today}% It is always \today, today,

\begin{abstract}
We obtain perturbative expressions for jet distributions using soft-collinear 
effective theory (SCET). By matching 
SCET onto QCD at high energy, tree level matrix elements and higher order 
virtual corrections can be reproduced in SCET. The resulting operators are then evolved
to lower scales, with additional operators being populated by required threshold 
matchings in the effective theory. We show that the renormalization group
 evolution and threshold
matchings reproduce the Sudakov factors and splitting functions of QCD, and that
the effective theory naturally combines QCD matrix elements and parton showers. 
The effective theory calculation is systematically improvable and any higher order perturbative
effects 
can be included by a well-defined procedure. 
\end{abstract}

\maketitle

%----------------------------------------------------------------

Calculating the distribution of jets in collider experiments is essential for
understanding the standard model and in looking for signals of new physics.
Unfortunately, improving the accuracy of a calculation is not
simply a matter of computing higher loop QCD diagrams
by brute force. These calculations produce large logarithms, which must
be resummed. Moreover, fixed order calculations are
only feasible for a modest number of partons.
For comparison with experimental data, 
however, it is desirable to obtain theoretical predictions for processes with
a much larger number of final-state particles. The traditional method is to produce the 
required distributions using splitting functions, which 
are derived in the limit of small transverse momentum.
In this limit, distributions factorize into lower order processes, multiplied
by the classical probabilities for particles to branch. 
The probability for no branching is given by a Sudakov factor, 
which is simply an integral over the splitting functions,
and sums the leading logarithmic (LL) terms.
Several programs use this method of
parton showers to 
generate distributions with an arbitrary number of final states and
are valid at LL~\cite{PSidea,PSprograms}.
Since the underlying assumption in parton showers is small transverse momentum, 
additional information from QCD matrix elements is required to properly describe
jets which are widely separated.

The issue of consistently matching fixed order QCD calculations with 
parton shower evolution has been the source of 
active research over the 
past several years~\cite{CKKW,FrixioneWebber,NagySoper}. 
One must avoid double counting between emissions described by QCD matrix elements
and those contained in the splitting functions, and cancel the 
infrared divergences between real and virtual QCD diagrams.
Most programs include some hard matrix element corrections~\cite{hardjet},
however there is no systematic way to improve their accuracy.

Fundamentally, the problem of jet distributions is one of scale separation. 
These scales
include the hard  scale $Q$ of the underlying interaction, 
and the scales $p_T^{(i)}$ for the relative 
transverse momenta of additional partons. 
Thus, it seems natural to
reformulate the problem in the language of effective field theory, where
scales are naturally separated. The appropriate effective theory for this 
problem is the soft-collinear effective theory (SCET)~\cite{SCET}. 
As we will show, the information of fixed order QCD
is contained in SCET by a matching calculation at the hard scale $Q$, 
while the summation of the large logarithms is achieved
by renormalization group (RG) evolution within SCET. Furthermore, in
contrast to other approaches,
the effective theory
allows us to systematically incorporate 
higher order effects by straightforward and well-defined calculations. 

The matching at the hard scale involves determining the Wilson 
coefficients of SCET
operators such that matrix elements in the effective
theory reproduce the full theory up to a fixed order in $\alpha_s$. Because
all the infrared physics of the full theory is reproduced in SCET, the
matching coefficients will be finite at any order in $\alpha_s$. 
To reproduce QCD with $N$ partons requires operators with 
up to $N$ fields in the effective theory. 
To calculate this matching, of course, requires knowledge of QCD matrix elements
with up to $N$ partons in the final state. 

After the matching, full QCD is no longer needed. 
The next step is to evolve  the operators in SCET
to  lower scales using the RG. 
This sums the large logarithms and, as we will
show, it reproduces the Sudakov factors at LL. 
Lowering the renormalization scale in SCET corresponds to 
lowering the value of $p_T$ accessible to fields in the effective theory. 
As the scale gets lowered below the $p_T$ of one of 
the partons in the
final state, the emission of this parton is no longer described by 
the interactions of SCET. 
To ensure that the theory still describes the same physics,
a threshold matching is required which gives rise to new operators with additional 
collinear fields. 
%This matching therefore populates
%operators with more than $N$ collinear fields. 
This matching involves SCET matrix elements for emission, which give
splitting functions in the collinear limit.

For more details, we need to understand the basic construction of SCET. We want 
to describe the long distance behavior of QCD matrix elements with several energetic partons 
in well-separated directions. To achieve this, SCET requires collinear fields for each of these directions,
 as well as soft fields they can interact with. Collinear fields $\xi_n$
 are labeled by a light-like direction 
$n^\mu$ and have energy much larger than the momentum transverse to $n^\mu$, $E \gg p^\perp$. 
Collinear fields in SCET appear in operators as jets $\chi_n$
which are fields wrapped in Wilson lines
\begin{eqnarray}
\chi_n = W_n \xi_n\,.
\end{eqnarray}
The Wilson lines ensure gauge invariance of the theory. For example, a two jet
operator is
\begin{eqnarray}
\cO_2 = {\bar \chi}_n \Gamma \chi_\bn
= {\bar \xi}_n W^\dagger_n \Gamma W_\nbar \chi_\nbar\,,
\end{eqnarray}
where $\Gamma$ denotes some Dirac structure. 
To correctly describe full QCD with up to $N$ jets in different directions requires
operators in SCET with up to $N$ collinear fields, each with their own label $n_i^\mu$.

Since the sum of a number of collinear momenta is still collinear, there
are interactions in SCET among particles collinear to each direction $n^\mu_i$.
Particles collinear to different directions do not interact directly, but can exchange
soft gluons.
The RG evolution in the theory, from higher to lower $p^\perp$,
resolves the transverse momentum of jets. It physically disassembles 
jets into their partonic constituents.

In this paper, we will study as an example the canonical process $e^+ e^- \to {\rm hadrons}$
(for more details, see~\cite{us2}), with  center-of-mass energy denoted by $Q$.
As mentioned above, the full matrix elements are matched onto SCET at the hard scale $Q$, 
while the RG evolution in the effective theory sums the logarithms. To begin,
we only include the 
QCD matrix element with 2 partons in the final state, which allows  us  to focus
on running. In this case, additional partons
arise from SCET emissions through the threshold matching. Thus, the results
will will only be valid in the limit $Q \gg p_T^{(1)} \gg p_T^{(2)} \gg \ldots\,,$
where $p_T^{(i)}$ denote the transverse momenta between particles in the final state. 
Note that this is the same limit in which parton showers are known to 
hold. Because the effective theory includes the RG evolution, it will 
correctly sum all the logarithms of ratios of these scales.

The matching condition is
\begin{equation}
\label{O2match}
\langle {\mathrm{QCD}} |q \bar{q} \rangle_Q=
\left[ \cC_2(\mu) \langle
  \cO_2 |q \bar{q} \rangle\,\right]_{\mu=Q},
\end{equation}
where $\langle \tmop{QCD} |q \bar{q} \rangle$ denotes the fixed order matrix element 
of the current $\qbar \Gamma q$. For particular final state momenta $q^\mu$, we define
$\xi_n | q \rangle=0$ if
$n^\mu$  is not aligned with $q^\mu$. That is, the matrix element is non-zero
only if $n^\mu = q^\mu/E_q$.
Then the matching is trivial and
\begin{eqnarray}
\cC_2(Q) = 1\, .
\end{eqnarray}

Having performed this matching, we can focus on the resummation 
of the large logarithms. This is achieved by running
the operator $\cO_2$ below the scale $Q$. The Wilson coefficients satisfy
the RG equation $d\log \cC_n(\mu)/d\log \mu = \gamma_n(\mu)$, which has the solution
\begin{equation}
\label{Cnrunning}
  \frac{\cC_n ( \mu_1 )}{\cC_n(\mu_2)} \equiv \Pi_n(\mu_2, \mu_1)=  \exp \left\{
  \int_{\mu_1}^{\mu_2} \frac{d \mu}{\mu} \gamma_n ( \mu ) \right\}\,.
\end{equation}
$\Pi_n$ is an evolution kernel which is similar to the Sudakov factor in traditional 
parton showers. 
The anomalous dimension $\gamma_2$ can be calculated by simple one-loop 
diagrams in SCET~\cite{manoharDIS}, giving
\begin{equation}
\label{gamma_2}
  \gamma_2 ( \mu ) = - \frac{\alpha_s ( \mu )}{\pi} C_F \left( \log \frac{-
  \mu^2}{Q^2} + \frac{3}{2} \right)\,,
\end{equation}
where $C_F = 4 / 3$. 
Note that it is only the logarithmic term in the anomalous dimension which gives 
rise to the LL resummation. The constant term $3/2$ gives rise to a subset of the 
subleading logarithms and can thus be omitted if only LL accuracy is desired. 
To obtain the complete NLL result, the coefficient
of the $\log \mu/Q$ term is required at two loops. Higher orders in the resummation
can straightforwardly be included by calculating the anomalous dimensions at even higher
loop order.

At a scale $p_T$ a third parton can be 
resolved in one of the jets. Below that scale the matrix element of $\cO_2$ with the 
three partons vanishes  and a threshold matching onto a new operator 
$\cO_3^{(2)}$ has to be performed (the superscript indicates that 
this operator arises through threshold matching).  The matching condition is 
\begin{equation}
\label{thresholdmatch}
  \left [\cC_2(\mu) \langle \cO_2 |q \bar{q} g\rangle \right]_{\mu = p_T+\epsilon} 
= \left[ \cC_3^{(2)}(\mu) \langle  \cO_3^{(2)} |q \bar{q} g\rangle\right]_{\mu = p_T-\epsilon}\,.
\end{equation}
Emissions in SCET can  be
collinear or soft, and the  collinear gluon can either be radiated from 
the collinear quark (anti-quark), $\bar{\xi}_n (\xi_\nbar)$ ,
 or it can be emitted by one of the Wilson lines, 
$W^\dagger_n (W_\nbar)$.
There are some ambiguities in calculating the matrix elements in SCET,
which will only affect higher order contributions if incorporated consistently. 
Here, we will take the additional gluon in the final state to be collinear, 
since soft gluons have much smaller $p^\perp$ than collinear gluons. We also use the fact that the 
emission of a collinear gluon from the collinear Wilson line will give 
rise to longitudinally polarized gluons (up to higher orders in $p_T/Q$). Thus, to 
calculate the matrix element of $\cO_2$ in Eq.~(\ref{thresholdmatch}) we 
include only the emission from the collinear fermions and keep only transverse gluon 
polarizations $\epsilon^\perp$.
With these conventions Eq.~(\ref{thresholdmatch}) is satisfied by $\cC_3^{(2)}(p_T) = \cC_2(p_T)$ and 
\begin{eqnarray}
\label{O32match}
\langle \cO_3^{(2)} | q \bar q g \rangle = g_s\bar \chi_{n_q} \left[ \frac{\epslash^\perp 
\frac{\bnslash_{\bar q}}{2} \Gamma}{\bn_{\bar q} \cdot (p_q+p_g)}
-\frac{\Gamma \frac{\bnslash_q}{2}
\epslash^\perp}{\bn_{q} \cdot (p_{\bar q}+p_g)} \right] \chi_{n_{\bar q}}\,.
\end{eqnarray}

Note that 
we choose to integrate out  emissions from both the quark and the anti-quark at the same
scale $p_T$. In general, the transverse momentum of the gluon with respect to the quark can differ from the 
one with respect to the anti-quark, and one could  choose to integrate out each emission at these
different scales. However, the difference between these scales
is subleading in $p_T/Q$, and thus beyond the order we are working. 

This procedure of matching and running is continued for 
each value of $p_T^{(i)}$. 
A closed form expression can be obtained for the anomalous dimension of
a general operator $\cO_n$ which contains
$n_q$ quark and $n_g$ gluon fields. At leading order
\begin{eqnarray}
\label{gamma_n}
\gamma_n = - \frac{\alpha_s(\mu)}{\pi}\log\frac{\mu^2}{Q^2}
\left[ \frac{n_f}{2} \, C_F + \frac{n_g}{2} \, C_A \right]\,,
\end{eqnarray}
and the solution to the RG equation is given by Eq.~(\ref{Cnrunning}). 
At leading log accuracy the evolution kernels for the Wilson coefficients are
directly related to the Sudakov factors for quarks and gluons, $\Delta_q$ and $\Delta_g$, 
used in traditional parton showers~\cite{PSprograms,Mrenna:2003if}
\begin{eqnarray}
\label{PiDeltaequiv}
\Pi_n(\mu_2,\mu_1) \stackrel{LL}{=}\Delta_q^{n_q/2}(\mu_2,\mu_1) \Delta_g^{n_g/2}(\mu_2,\mu_1)\,.
\end{eqnarray}

The threshold matching can be continued for additional emissions.
In general, the rule for quark splitting in SCET can be written as
\begin{eqnarray}
\label{fermionsplit}
\chi_n \to g_s \frac{\nslash}{2} \frac{\pm \epslash^\perp}{n\tdot (p_{\bar q}+p_g)}  \chi_{n'}\,,
\end{eqnarray}
where  the $+(-)$ is for fermions (antifermions). A similar equation can be obtained
for gluons splitting into two gluons or two fermions. 
New operators $\cO_m^{(2)}$ are thus obtained by replacing each collinear field 
in $\cO_{m-1}^{(2)}$ with 
Eq.~(\ref{fermionsplit}) and summing over all contributions. 
Note that the operators $\cO_m^{(2)}$ are sums over many different terms, each 
giving different splitting histories, just as 
the operator $\cO_3^{(2)}$ in Eq.~(\ref{O32match}) is the sum of splittings
of the quark and the anti-quark. 

This sequence of threshold matching and running is repeated until a low scale 
$\mu = \mu_0$ is reached, where non-perturbative physics becomes important. At that scale 
the amplitude of $e^+ e^- \to {\rm hadrons}$ is
\begin{eqnarray}
\label{APS}
\langle \mathrm{SCET} \rangle_{\mu_0}  &=& 
\sum_m \cC_m^{(2)}(\mu_0) \,\langle \cO_m^{(2)} \rangle_{\mu_0} \, ,
\end{eqnarray}
where the evolution of the operators between the various threshold matching scales is captured 
in $\cC_m^{(2)}(\mu_0)$
\begin{eqnarray}
\label{Pigeneral}
\cC_m^{(2)}(\mu_0) = \cC_2(Q) \Pi_2(Q,p_T^{(1)})\,\cdots \, 
 \Pi_{m-1}(p_T^{(m-3)}\!\!\!,\mu_0)\,.
\end{eqnarray}

To compute differential cross sections, we need to square this amplitude and sum over final 
spins of the fermions and transverse polarizations of the gluons. 
Because the operator $\cO_m^{(2)}$ is obtained from the operator $\cO_2$ by iterating
Eq.~(\ref{fermionsplit}), we can simplify the 
form of $|\cO_m^{(2)}|^2$. To leading order in $p_T/Q$, the result is
%Eq.~(\ref{fermionsplit}) implies
\begin{eqnarray}
\label{MEsq}
&&\sum_{\rm spins,pols} \left|\cO_m^{(2)}\right|^2  = 
\sum_{\rm spins,pols} \left|\cO_{m-1}^{(2)}\right|^2 \frac{1}{p_T^2}P(z)\,,
\end{eqnarray}
where $P(z)$ is a splitting function, and the variable $z$ denotes the 
ratio of the energy of the emitted particle to the energy of the particle it was emitted from. 
Each emission of an additional particle gives rise to a splitting function 
after the matrix element is squared. Using 
Eq.~(\ref{MEsq}) together with Eq.~(\ref{PiDeltaequiv}), 
the cross section obtained from the 
amplitude $\langle {\rm SCET} \rangle$ 
given in Eq.~(\ref{APS}) reduces to the cross section of $e^+ e^- \to q \bar q$ scattering, 
multiplied by products of splitting functions and Sudakov factors. Thus, SCET is equivalent
to the traditional parton shower at leading order.

Now that we have seen how SCET sums the large logarithms and showers
additional collinear particles, we can go back and incorporate the QCD matrix element with 
an additional gluon in the 
final state. Thus, we should include $\cO_2$ and an operator 
$\cO_3$ in the matching at $\mu = Q$. 
%(A superscript is added to distinguish $\cO_3^{(3)}$ from the operator $\cO_3$ which arose from 
%the threshold matching at the scale $p_T$.) 
This allows us to describe the differential rate with three
well-separated jets,
for any value of $p_T^{(1)}$, and
will improve our previous results such that the resulting distributions are valid in 
the limit $Q, p_T^{(1)} \gg p_T^{(2)} \gg p_T^{(3)} \ldots\,.$ In order to obtain 
the correct normalization of the total cross section at ${\mathcal O}(\alpha_s)$, the 
one loop matching onto the operator $\cO_2$ is required at the hard scale as well. 

To match onto $\cO_3$ we calculate matrix elements in QCD and SCET
with three partons in the final state
\begin{equation}
\label{hardmatch}
  \langle \tmop{QCD} |q \bar{q} g \rangle = \cC_2 \langle
  \cO_2 |q \bar{q} g \rangle + \cC_3 \langle \cO_3 |q
  \bar{q} g \rangle\,.
\end{equation}
The operator $\cO_2$ and its Wilson coefficient $\cC_2$ at order 
$\alpha_s^0$
have already been determined
from the matching with two partons. With our convention for emissions
\begin{eqnarray}
\label{O33}
\langle \cO_3 | q \bar q g \rangle = 
 g_s \bar \chi_{n_q} \left[ \frac{\epslash^\perp \frac{\nslash_{\bar q}}{2} \Gamma}
{\nslash_{\bar q} \tdot (p_q + p_g)}
 -\frac{\Gamma \frac{\nslash_q}{2}  \epslash^\perp}{\nslash_q\tdot(p_{\bar q} + p_g)} \right] 
\chi_{n_{\bar q}}\,.
\end{eqnarray}
and then Eq.~(\ref{hardmatch}) is satisfied for $\cC_3(Q) = 1$.
This matrix element vanishes in the limit that the gluon is soft or is collinear
to either the quark or the anti-quark. This is again due to the fact that SCET
reproduces the IR of QCD.

To obtain $\cC_2$ to order $\alpha_s$ we again use Eq.~(\ref{O2match}), but 
now calculate the matrix elements in QCD and SCET at one-loop. Since the IR divergences
of QCD are reproduced in SCET, the Wilson coefficient is finite~\cite{manoharDIS}
\begin{eqnarray}
\cC_2(Q) = 1 - \frac{\alpha_s C_F}{4 \pi} \left( 8 - \frac{7\pi^2}{6} + 3 \pi i \right)\,.
\end{eqnarray}
Even higher order corrections to these Wilson coefficients can be included
by performing the matching at higher loop order.

After the matching at $\mu=Q$, the operators are run down to a low scale.
The running of $\cO_3$ is identical to the running of $\cO_3^{(2)}$, since
running in SCET depends only on the field content of an operator, not on
its tensor structure. We run both $\cO_2$ and $\cO_3$ down to
the scale $\mu\sim p_T$, at which point we match $\cO_2$ onto $\cO_3^{(2)}$ as
before. At $\mu \sim p_T$
\begin{eqnarray}
\label{A3>Q0}
\langle {\rm SCET}^{(3)} \rangle_{p_T} = \cC_2\,\Pi_2\,\langle \cO_3^{(2)}\rangle
+ \cC_3 \,\Pi_3\,\langle \cO_3 \rangle \,,
\end{eqnarray}
After that, the sequence of running and matching continues as before,
but now with two series of operators: $\cO_m^{(2)}$, populated by branchings descended
from $\cO_2$, and $\cO_m^{(3)}$, which are descended from $\cO_3$.
The final result at the scale $\mu_0$ is
\begin{eqnarray}
\langle {\mathrm{SCET}}^{(3)} \rangle_{\mu_0} = 
\Pi_3( p_T^{(1)},p_T^{(2)})\cdots\Pi_{m-1}(p_T^{(m-3)},\mu_0) \times \quad
\label{scet3}\\
\left[\cC_2(Q) \Pi_2(Q,p_T^{(1)}) \langle \cO_m^{(2)} \rangle
+\cC_3(Q) \Pi_3(Q,p_T^{(1)}) \langle \cO_m^{(3)}\rangle \right] \, .
\nonumber
\end{eqnarray}
Note that this is very similar to what we
had in Eqs.~\eqref{APS} and~\eqref{Pigeneral}. 
But it is correct to $\cO(\alpha_s)$ and for values
of $p_T$ that satisfy $Q, p_T^{(1)} \gg p_T^{(2)} \gg p_T^{(3)} \ldots\,$. It reproduces
the three jet differential cross section at NLO, but also sums all leading logarithms.

\begin{figure}[t!]
\begin{center}
 \includegraphics [width=0.50\textwidth,clip]{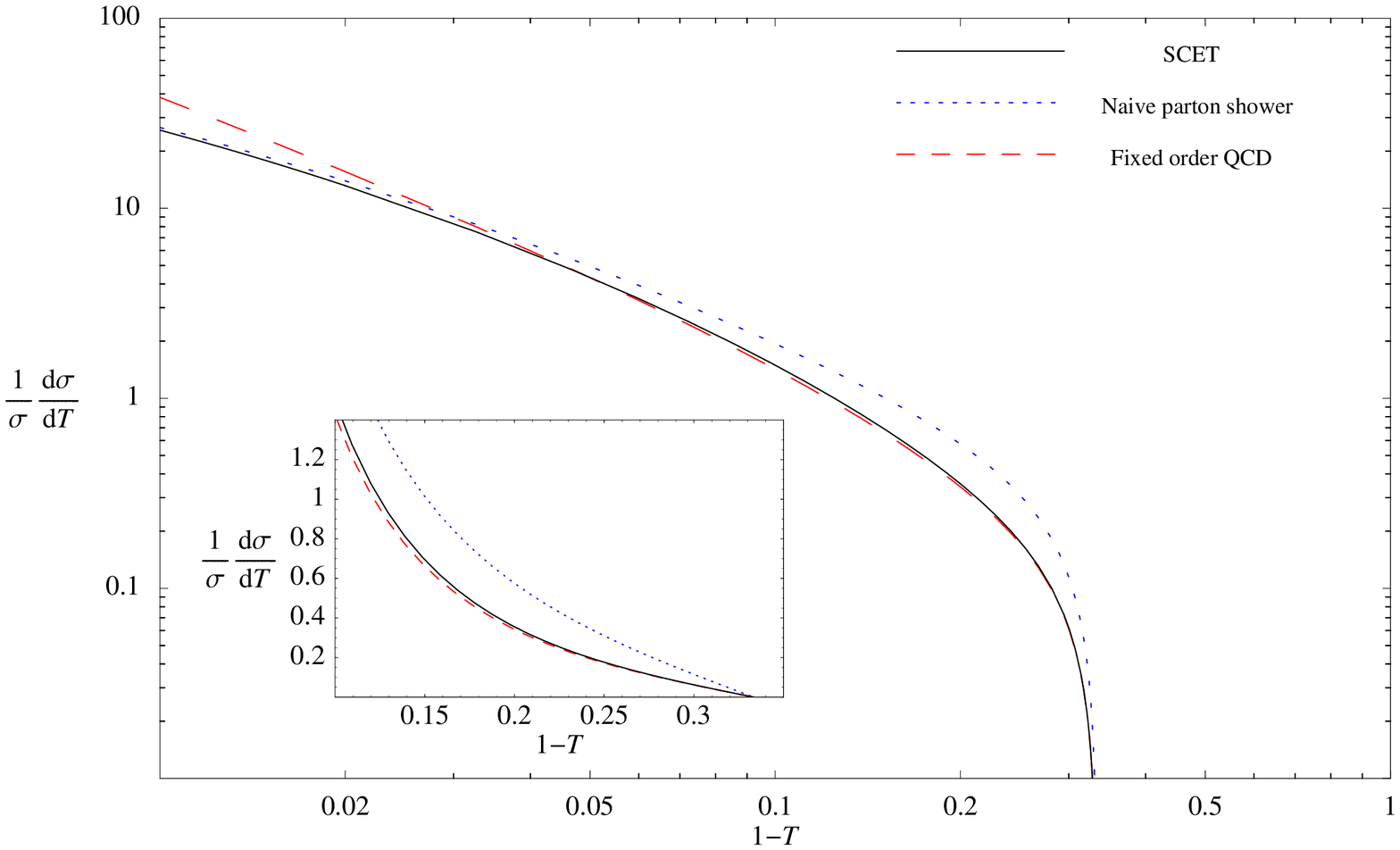}
 \caption{Thrust distribution in 3-jet events, calculated as described in the text.
Note how the SCET result
reduces
to QCD at small thrust (large $p_T$) and to the 
parton shower at large thrust (small $p_T$).} 
\label{f1}
\end{center}
\end{figure}

The SCET result interpolates between fixed 
order QCD and a parton shower. To demonstrate this, it is enough
construct observables based on 3-jet final states. Since we only 
need information about the first branching we should evaluate 
the matrix elements at a  scale of order $p_T$, and the result is given by Eq.~(\ref{A3>Q0}). 
It is not hard to see that Eq.~(\ref{A3>Q0}) has the right limits.  
For $p_T \ll Q$,  SCET is an excellent approximation to QCD and 
the matrix element of $\cO_3$, which represents the difference between
QCD and SCET, vanishes. 
Also, $|\langle \cO_3^{(2)}|q\qbar g\rangle|^2$ reduces to a splitting function in this limit,
%
%Furthermore, the extra emission in $O_3$ reduces to 
%the splitting function in the same limit,  
as indicated in Eq.~(\ref{MEsq}), and $\Pi_2$
reproduces the LL Sudakov factor, as shown in Eq.~(\ref{PiDeltaequiv}). 
Thus, for $p_T\ll Q$, SCET reduces to the product of a Sudakov factor 
and a splitting function, as in a parton shower. In contrast, for $p_T \sim Q$, SCET$^{(3)}$
is very close to QCD, since SCET has not evolved far from the scale where we matched
it to QCD exactly.
 Given the results in these two limits, the effective theory will smoothly interpolate
between QCD and the parton shower. To see this, in Fig.~\ref{f1} we show
the differential thrust distribution for $Q=91$ GeV. For three partons, 
$T = {\rm max}\{E_q,E_\qbar,E_g\}/2Q$.
The SCET result is given by squaring $\eqref{A3>Q0}$
and integrating over the remaining phase space, while for QCD we use the leading tree
level differential cross section. For the 
parton shower, we use the splitting functions $1/(p_q+p_g)^2 P(z)$, multiplied by the LL Sudakov factor. We have chosen the value of the scale  $\mu\sim p_T$ such that the maximum value of the Wilson coefficients and Sudakov factors is unity.

We have shown that SCET reproduces QCD at NLO, including full parton showering.
There are
many ways the results presented here can be improved: 
1) Include more operators at $\mu = Q$.
2) Include the matching at $\mu = Q$ to higher order in $\alpha_s$.
3) Include the running of the operators at subleading order.
4) Include the threshold matching in SCET at higher order.
5) Include power corrections from the SCET expansion.
It is important to bear in mind that all these improvements are straightforward calculations
in the effective theory. For example, to include more operators at $\mu = Q$ simply requires
additional tree level calculations in QCD and the effective theory, while to go beyond NLO only 
requires the calculation of well-defined loop diagrams in QCD and SCET. No new 
formalism needs to be developed to achieve any of these results. 
Thus, SCET is a
powerful tool for improving the theoretical understanding of jet distributions.
It provides a convenient systematically improvable
framework for performing higher loop calculations, and should be straightforward
to implement in an event generator, allowing direct comparison to data.

We would like to thank Steve Mrenna, Peter Richardson and Peter Skands for comments on the manuscript. This work was supported by the DOE under Contract DE-AC03-76SF00098.
% and DE-FG03-91ER-40676. 

%----------------------------------------------------------------

\end{document}